\documentclass[PRX,reprint,amsmath,amssymb,superscriptaddress,longbibliography]{revtex4-1}

\usepackage{graphicx}
\usepackage{dcolumn}
\usepackage{bm}
\usepackage{hyperref}
\usepackage{breakurl}
\usepackage{multirow}
\usepackage{enumitem}
\setcitestyle{super} %

\begin{document}\title{Topological phase transition induced extreme magnetoresistance in TaSb$_{2}$}

\author{Zhen Wang}
      \thanks{Equal contributions}
      \affiliation{Department of Physics, Zhejiang University, Hangzhou 310027, P. R. China}
       \affiliation{State Key Lab of Silicon Materials, Zhejiang University, Hangzhou 310027, P. R. China}

\author{Yupeng Li}
      \thanks{Equal contributions}
      \affiliation{Department of Physics, Zhejiang University, Hangzhou 310027, P. R. China}

\author{Yunhao Lu}
      \affiliation{State Key Lab of Silicon Materials, Zhejiang University, Hangzhou 310027, P. R. China}

\author{Zhi-Xuan Shen}
      \affiliation{Department of Physics, Zhejiang University, Hangzhou 310027, P. R. China}

\author{Feng Sheng}
      \affiliation{Department of Physics, Zhejiang University, Hangzhou 310027, P. R. China}

\author{Chunmu Feng}
      \affiliation{Department of Physics, Zhejiang University, Hangzhou 310027, P. R. China}

\author{Yi Zheng}
      \email{phyzhengyi@zju.edu.cn}
      \affiliation{Department of Physics, Zhejiang University, Hangzhou 310027, P. R. China}
      \affiliation{Zhejiang California International NanoSystems Institute, Zhejiang University, Hangzhou 310058, P. R. China}
            \affiliation{Collaborative Innovation Centre of Advanced Microstructures, Nanjing 210093, P. R. China}

\author{Zhu-An Xu}
   \email{zhuan@zju.edu.cn}
      \affiliation{Department of Physics, Zhejiang University, Hangzhou 310027, P. R. China}
      \affiliation{State Key Lab of Silicon Materials, Zhejiang University, Hangzhou 310027, P. R. China}
      \affiliation{Zhejiang California International NanoSystems Institute, Zhejiang University, Hangzhou 310058, P. R. China}
            \affiliation{Collaborative Innovation Centre of Advanced Microstructures, Nanjing 210093, P. R. China}

\date{\today}

\begin{abstract}

We report extremely large positive magnetoresistance of 1.72 million percent in single crystal TaSb$_{2}$ at moderate conditions of 1.5 K and 15 T. The quadratic growth of magnetoresistance (MR $\propto\,B^{1.96}$) is not saturating up to 15 T, a manifestation of nearly perfect compensation with $<0.1\%$ mismatch between electron and hole pockets in this semimetal. The compensation mechanism is confirmed by temperature-dependent MR, Hall and thermoelectric coefficients of Nernst and Seebeck, revealing two pronounced Fermi surface reconstruction processes without spontaneous symmetry breaking, \textit{i.e.} Lifshitz transitions, at around 20 K and 60 K, respectively. Using quantum oscillations of magnetoresistance and magnetic susceptibility, supported by density-functional theory calculations, we determined that the main hole Fermi surface of TaSb$_{2}$ forms a unique shoulder structure along the $F-L$ line. The flat band top of this shoulder pocket is just a few meV above the Fermi level, leading to the observed topological phase transition at 20 K when the shoulder pocket disappears. Further increase in temperature pushes the Fermi level to the band top of the main hole pocket, induced the second Lifshitz transition at 60 K when hole pocket vanishes completely.
\end{abstract}

\maketitle

\section{Introduction}
Topological classifications of materials beyond the conventional band theory is one of the revolutionary advances in condensed-matter physics, as started by the discovery of topological insulators (TIs) \cite{TI_PRL_Kane,TI_Science06_ZSC,TI_RMP10_Kane}. Searching for new archetype of topological materials has intrigued enormous research efforts, and has witnessed the birth of Dirac Semimetals (DSMs) \cite{DSM_PRLtheory_Kane,Cd3As2_PRBtheory_LMR,Cd3As2_NPOng_NatMat15}, Weyl semimetals (WSMs) \cite{WSMWanXG_PRB,WSMDaiX_PRX,NCHasan_WSMTheory}, nodal-line states \cite{PbTaSe2_Hasan_Nodal}, and numerous theoretical proposals for various topological phases \cite{QuatumPhases_Balents_AR}.

Intriguingly, many topological semimetals are characterized by extremely large magnetoresistance (XMR). The linear XMR in DSM Cd$_{3}$As$_{2}$ has been attributed to the lifting of protection mechanism by the external magnetic field \cite{Cd3As2_NPOng_NatMat15}, however, Narayanan \textit{et al.} argued that such linear behavior is disorder related \cite{Cd3As2_PRLXMR_Coldea}. For the inversion symmetry broken TaAs-family WSMs, XMR is also non-saturating, but quasi-linear in field dependence \cite{TaAs_arXiv_Jia,TaAs_PRX_NMR,NbAs_jpcm,NbP_NaturePhy,NbP_arXivWZ}, which is explained by electron-hole ($e$-$h$) compensation between the co-existing trivial and WSM pockets \cite{TaAs_arXiv_Jia,NbP_NaturePhy,NbP_arXivWZ}. In contrast, similar $e$-$h$ compensation mechanism leads to distinct quadratic MR in WTe$_{2}$ \cite{XMR_WTe2NPOng_Nature14}. Resonant compensation with highly matched $e$ and $h$ pockets in WTe$_{2}$ is widely accepted to be responsible for the experimental observation, as supported by angle-resolved photoemission spectroscopy (ARPES) \cite{Aupes_PRL_Valla} and angle-dependent quantum oscillations \cite{FS_PRL_ZhuZW}. However, Jiang \textit{et al.} proposed that strong spin-orbital coupling plays a non negligible role in the quadratic XMR using circular polarized ARPES \cite{SOC_Aupes_PRL_FengDL}, and Rhodes \textit{et al.} showed that the Fermi surface of WTe$_{2}$ is significantly modified by magnetic field due to the Zeeman coupling \cite{Zeeman_PRB_Balicas}. Very recently, XMR is also reported in rocksalt structured LaSb \cite{LaSb_Cava_XMR}, which is claimed to be a partially compensated semimetal with dominant electron concentration of $\sim 1\times10^{20}$ cm$^{-3}$ at low temperatures. The XMR of LaSb with high residual resistivity ratio (RRR $>$ 800) is quadratic-like, but becomes saturating for lower quality samples \cite{LaSb_Cava_XMR}.

Noticeably, Cd$_{3}$As$_{2}$, the TaAs series, and WTe$_{2}$ \cite{WSMIIWTe2_Bernevig_Nature15} all have lattice symmetry protected topological nodes in the bulk, while LaSb may host topological surface states as a TI candidate with a band gap of 10 meV \cite{LaSbTI_LinHsin_arXiv}. In the present study, we report exotic quadratic XMR in monoclinic TaSb$_{2}$, which strictly follows the the $B^{1.96}$ dependence and reaches an extremely large MR of 1.72 million percent at 1.5 K and 15 T. Using temperature-dependent MR, Hall and thermoelectric coefficient measurements, we found that TaSb$_{2}$ has perfectly compensated $e$-$h$ pockets at 1.5 K, with a mismatch between electron density $n_{e}$ and hole density $n_{h}$ less than 0.1$\%$, in contrast to 4$\%$ in WTe$_{2}$. By increasing temperatures, TaSb$_{2}$ undergoes two subsequent Lifshitz transitions around 20 K and 60 K, respectively.  Using temperature-dependent quantum oscillations and first-principle density-functional theory calculations, we determine the physical origin of these two topological phase transitions, which are rooted in the unique shoulder structure of the main hole pocket along the $L-I$ direction. Our study suggests that the Fermi surface topology may play an important role in the XMR phenomena of various topological semimetals.  

\begin{figure*}[!thb]
\begin{center}
\includegraphics[width=7in]{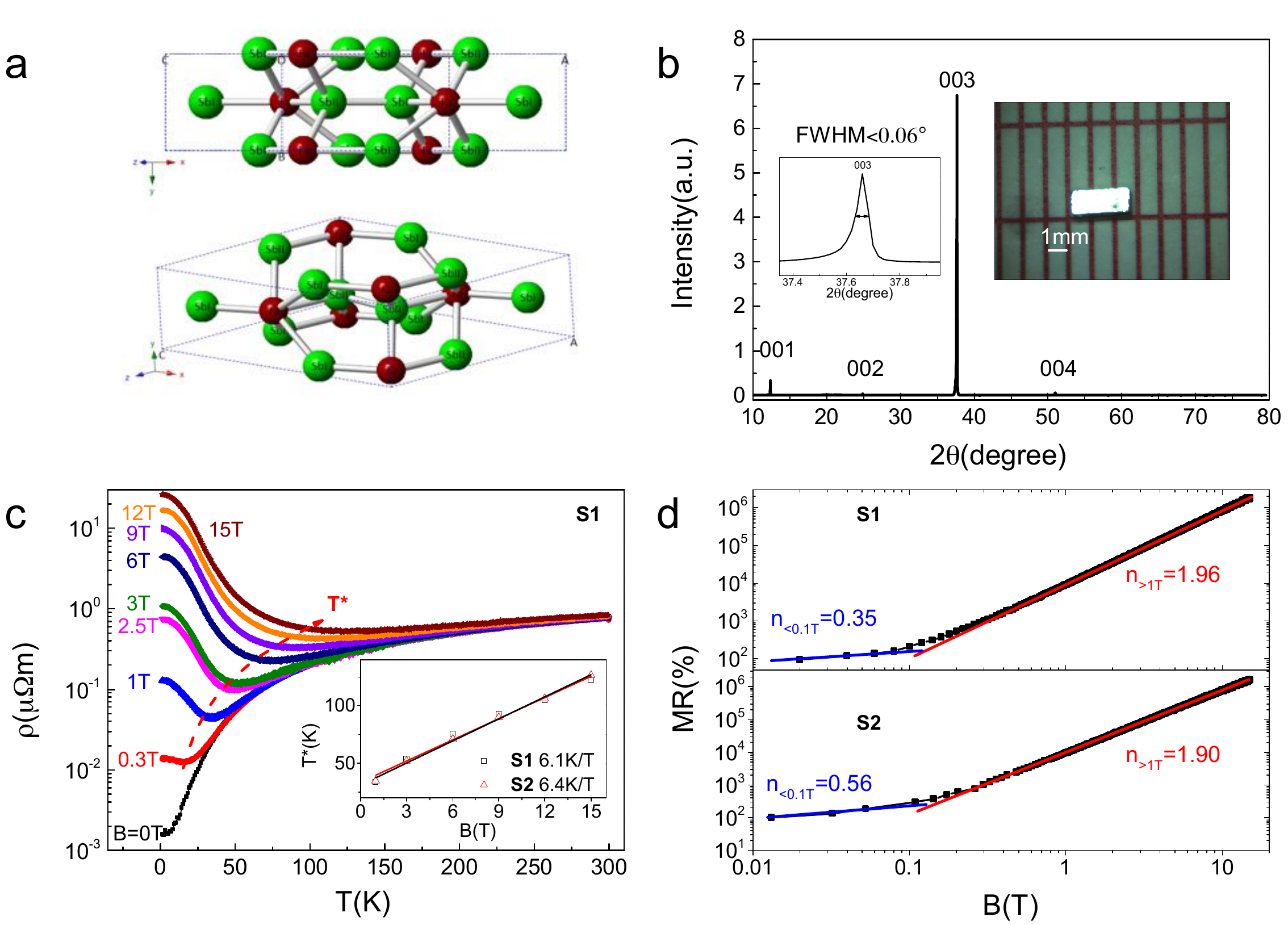}
\end{center}
\caption{\label{Figure1} Crystal structure and XMR of TaSb$_{2}$. (a) Monoclinic lattice of TaSb$_{2}$. There are two inequivalent positions for Sb atoms, labeled as Sb$_{1}$ and Sb$_{2}$ respectively. (b) XRD peaks of single-crystal TaSb$_{2}$. The incident angle is defined with reference to the (001) surface. (c) $\rho_{xx}$ vs T with fixed B, showing a transition from metallic to insulating crossing T$^{\ast}$. (d) Logarithmic plot of MR vs B at 1.5 K, showing strict B$^{m}$ growth up to 15 T.}
\end{figure*}

\section{Experimental}
Single crystal TaSb$_{2}$ was synthesized by two-step vapor transport technique using iodine as the transport agent. A stoichiometric mixture of high purity powders of Ta (99.99\%) and Sb (99.999\%) was thoroughly ground and pressed into pellets in an argon-filled glove box. The pellets were then sealed in an evacuated quartz tube and heated at 1023 K for 2 days. The resulting polycrystalline TaSb$_{2}$ pellets were ground into powder again, and mixed with iodine ($\sim$13 mg/ml in concentration) in a sealed quartz tube. Then, the tube was placed in a two-zone furnace with a temperature gradient of 50 K from 1273 K to 1223 K for 7 days. Shining single crystals with typical dimensions of $3\, mm \times1\, mm\times0.5 \, mm$ were got after the furnace was cooled down to room temperature naturally.

The crystal structure was characterized by X-ray diffraction (XRD) using a PANalytical X'€™Pert MRD diffractometer with Cu K$_{\alpha}$ radiation and a graphite monochromator. The chemical compositions of Ta:Sb=1:2 were determined by energy dispersion X-ray spectroscopy (EDX), showing no iodine residual in the single crystals. All electric transport measurements were carried out in an Oxford-15T cryostat with a He4 probe in the standard four-point contact or Hall-bar configurations, using Keithley 2400 source-measure meters and 2182A nanovoltmeters. The thermoelectric properties were measured by the steady-state technique. The magnetic field was applied along the $c$ axis while a temperature gradient about 0.5 K/mm was applied along the $b$ axis. The temperature differences were determined by the differential method using a pair of type E thermocouples.

The density-functional theory (DFT) calculations were carried out using the projector augmented wave method \cite{DFT_Blochl_PRB94}, as implemented in the Vienna \textit{ab initio} simulation package (VASP) \cite{VASP_Kresse_PRB93,VASP_Kresse_PRB96}. The exchange-correlation potential including spin-orbit coupling were applied by the generalized gradient approximation (GGA) developed by Perdew and Wang \cite{GGA_Perdew_PRL96}. The plane-wave cutoff energy was set to be 400 eV and k-point sampling based on the Monkhorst-Pack scheme \cite{MPscheme_Monkhorst_PRB76} was performed to ensure that the total energy is converged within 0.002 eV per unitcell. The structures were optimized until the remanent Hellmann-Feynman force on each ion is less than 0.01 eV/{\AA}.

\section{Results}

\begin{figure*}[!thb]
\begin{center}
\includegraphics[width=7in]{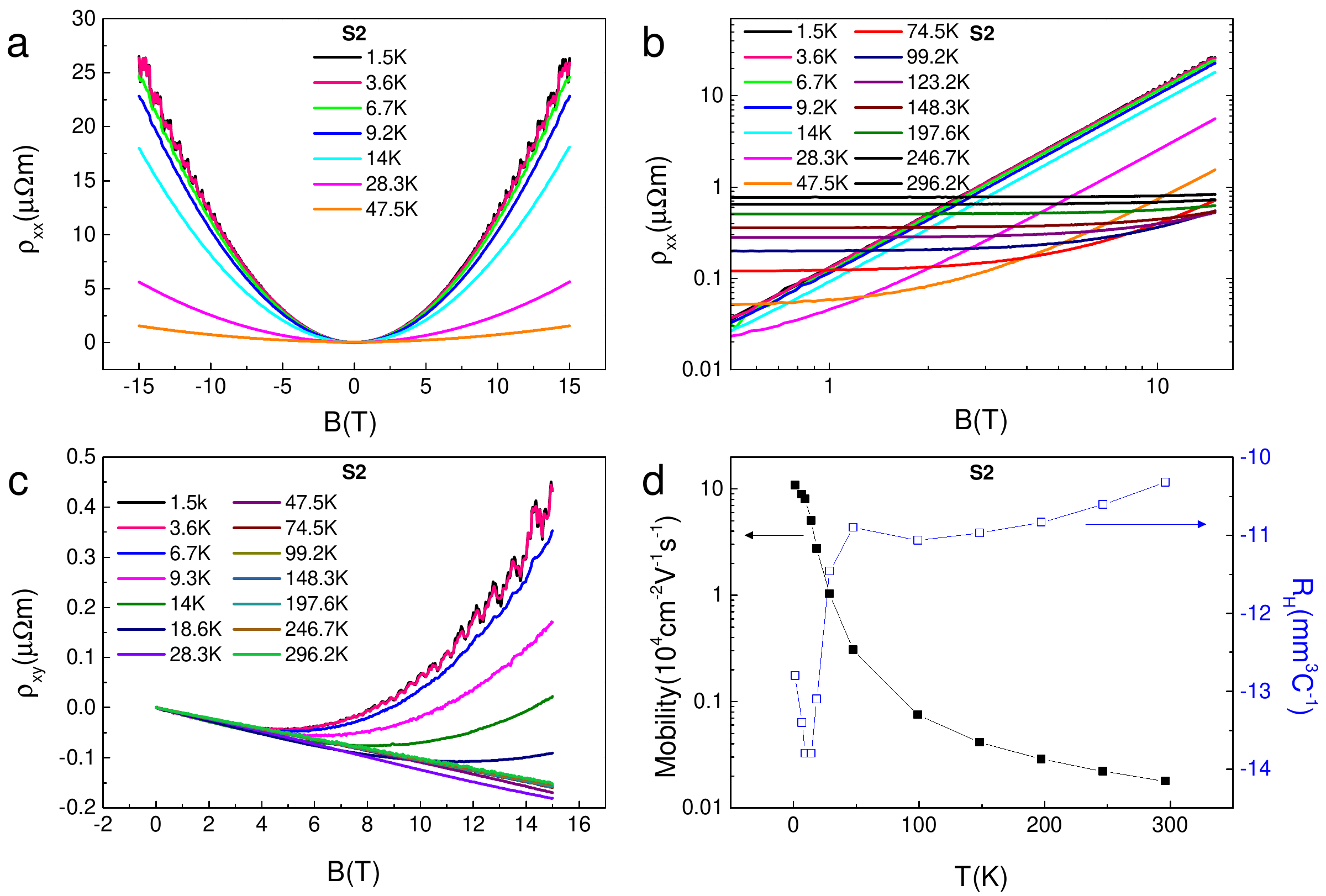}
\end{center}
\caption{\label{Figure2} Two subsequent Lifshitz phase transitions in TaSb$_{2}$ revealed by T-dependent MR and Hall. (a) The evolution of $\rho_{xx}$ vs B as a function of T. (b) The logarithmic plots of $\rho_{xx}$ vs B show the deviation of MR from $\bar{\mu} B^{1.96}$ when T approaches 60 K. (c) T-dependent Hall signals, showing a distinct Lifshitz transition at around 20 K. (d) Charge carrier mobility and R$_{H}$ calculated by the single-band model, revealing two Lifshitz transition temperatures at 20 K and 60 K respectively. The data were extracted by the linear fitting of $\rho_{xy}$ below 2 T.}
\end{figure*}

Figure 1a shows the crystal structure of TaSb$_{2}$, which is monoclinic with the space group of $C12/m1$ (No. 12). The upper panel of Fig. \ref{Figure1}a highlights the mirror symmetry, while the bottom panel clearly shows the inversion symmetry of the lattice. The XRD patterns indicate the high quality of our TaSb$_{2}$ single crystals, which shows the dominant [003] peak with very narrow full width at half maximum, and no splitting of peaks. The lattice parameters extracted from powder XRD are $a=10.225(1)$ {\AA}, $b=3.646(1)$ {\AA} and $c=8.295(1)$ {\AA} respectively. With zero magnetic field (B), the resistivity ($\rho_{xx}$) of TaSb$_{2}$ is metallic (the black line in Fig. \ref{Figure1}c). Once B is applied along the $c$ axis, \textit{i.e.} the [001] direction, TaSb$_{2}$ changes from metallic to insulating behavior at a temperature turning point ($T^{\ast}$), which is defined as the resistivity minimum. $T^{\ast}$ grows linearly as a function of B setpoint with a slope of 6.4 K/T, which is significantly higher than 4.4 K/T reported in WTe$_{2}$ \cite{XMR_WTe2NPOng_Nature14}. Like many other topological semimetals, the XMR growth in TaSb$_{2}$ is very sensitive to RRR. As shown in the upper panel of Fig. \ref{Figure1}d, for sample 1 (S1) with RRR=530, the quadratic growth is nearly ideal with $m$=1.96 in the logarithmic plot of MR vs $B^{m}$. For sample 2 (S2) with lower RRR=320, $m$ is reduced to 1.9. Similar to WTe$_{2}$, the XMR of TaSb$_{2}$ shows a characteristic ``activation'' field of 0.1 T, below which $m$ is smaller than 1. Despite the difference in $m$, the XMR growth in both samples does not show any tangible deviation from $\mathrm{B}^{m}$. This is in contrast with WTe$_{2}$, which shows decreasing $m$ when B exceeding 12 T, even for the highest quality samples with RRR $>$ 1200 \cite{XMR_WTe2NPOng_Nature14,FS_PRL_ZhuZW}. The faster growth in T$^{\ast}$ together with the non-saturating $m$ in TaSb$_{2}$ produce an extraordinary XMR of 1.72 million percent at 1.5 K and 15 T, among the highest records in reported literatures \cite{XMR_WTe2NPOng_Nature14,NbP_NaturePhy,NbP_arXivWZ,LaSb_Cava_XMR}. As a direct comparison of resonant compensated semimetals, WTe$_{2}$ has 4\% mismatch between $n_{e}$ and $n_{h}$ \cite{FS_PRL_ZhuZW}, thus, the even higher XMR with constant $m$ imply an even higher degree of resonant $e$-$h$ compensation in TaSb$_{2}$.

To get insight into the XMR mechanism, we have studied the temperature(\textit{T})-dependent $\rho_{xx}(B)$ of TaSb$_{2}$. As summarized in Figure \ref{Figure2}, below 60 K, the $\bar{\mu} B^{1.96}$ behavior of S2 is strictly reproduced at high B, although the XMR coefficient $\bar{\mu}$ decreases and the ``activation'' field increases when the sample is warmed up. The trend is more revealing using the logarithmic plot of the $\rho_{xx}(B)$ curves, which clearly shows the deviation of $\rho_{xx}(B)$ from $B^{1.96}$ when \textit{T} exceeds 60 K. Such a transition is also manifested in Hall signals, but with a distinct transition \textit{T} of 20 K. As shown in Fig. \ref{Figure2}c, below 20 K, $\rho_{xy}(B)$ shows pronounced sign reversal from negative to positive. Using the two-band model \cite{HallinMetal_Colin}, we found that the ``U''-shaped $\rho_{xy}(B)$ at 1.5 K is due to nearly perfect compensation of $n_{e}=4.627\times10^{19}\, \mathrm{cm}^{-3}$ and $n_{h}=4.628\times10^{19}\, \mathrm{cm}^{-3}$ ($<0.1\%$ mismatch), while the difference between $\mu_{e}$ and $\mu_{h}$ ($\sim10\%$) only plays a minor role in determining the curve shape [see the Supplemental Information (SI)]. Such high-precision resonant compensation persists up to 18.6 K. However, once \textit{T} crosses the critical point of 20 K, the Hall signals become linear with a negative coefficient  over the whole field range, which makes the two-band analysis rather tricky and subjective. It is noteworthy that, below 2 T, the $\rho_{xy}(B)$ curves are all linear with negative Hall coefficients, indicating the dominance of electron carriers even in the compensation regime. Thus, by assuming a single band conduction over the whole temperature range, we can extract important informations on the emergence of hole pockets and the resonant compensation, both effectively reducing the negative slope of $\rho_{xy}(B)$. Indeed, we have observed two transition temperatures of $\sim$ 60 K and $\sim$ 20 K, respectively in the single-band Hall coefficient R$_{H}$ (Fig. \ref{Figure2}d).

\begin{figure}[!thb]
\begin{center}
\includegraphics[width=3.5in]{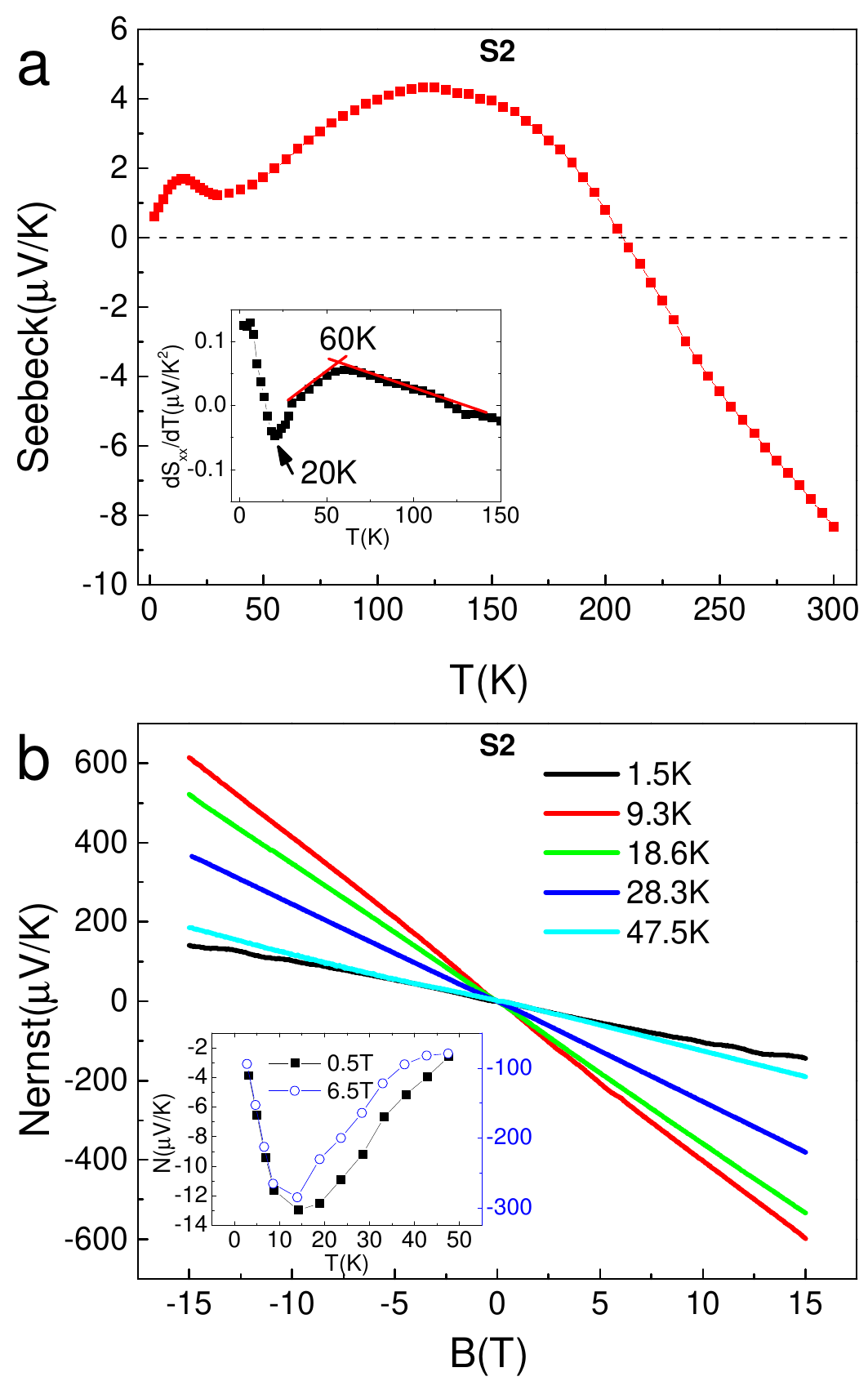}
\end{center}
\caption{\label{Figure3} (a) Lifshitz phase transition induced anomaly in the Seebeck coefficient. Two transition temperatures of 20 K and 60 K can be determined from $d \mathrm{S}_{xx}(\mathrm{T})/d\mathrm{T}$. (b) Linear Nernst vs B persists up to 47.5 K, implying the main $h$ pocket in TaSb$_{2}$ is comparable in size to the $e$ pockets above the first transition point of 20 K.}
\end{figure}

To understand the discrepancy in transition temperatures as inferred differently by \textit{T}-dependent $\rho_{xx}(B)$ and $\rho_{xy}(B)$ respectively, we turn to thermoelectric coefficients of Seebeck and Nernst effect, which provide indispensable information in understanding resonant compensated systems \cite{FS_PRL_ZhuZW}. For WTe$_{2}$, Wu \textit{et al} also shows that the \textit{T} driven deviation from resonant compensation is essentially a continuous Lifshitz transition \cite{Lifshitz_PRL_Kaminski}. Such topological Fermi surface reconstruction induces an anomaly in the Seebeck coefficient, close to the transition point \cite{Lifshitz_PRL_Kaminski}. In Figure \ref{Figure3}a, we show the \textit{T}-dependent Seebeck coefficient [$\mathrm{S}_{xx}(T)$], which switches the sign from negative to positive at around 205 K, an indication of the emergence of hole pocket at low temperatures. By taking the derivative of $\mathrm{S}_{xx}(T)$, we can clearly see a local maximum at 20 K, agreeing with the Hall results. In Ref. \cite{Lifshitz_PRL_Kaminski}, the Lifshitz transition is defined as the turning point in the slope of $d \mathrm{S}_{xx}(T)/dT$, representing the temperature where the hole pocket disappear completely \cite{Lifshitz_PRL_Kaminski}. In TaSb$_{2}$, we can find the same type of slope turning point at 60 K, consistent with the transition temperature suggested by the resistivity curves. The existence of a second Lifshitz transition at 60 K implies that the topological phase transition at 20 K does not lead to the vanish of the hole populations. Indeed, above 20 K, the hole pocket is still comparable to the electron one, as the linear dependence of Nernst ($\mathrm{S}_{xy}$) on magnetic field \cite{FS_PRL_ZhuZW} is evident at 47.5 K.

Using T-dependent Shubnikov-de Haas (SdH) and magnetic susceptibility oscillations, combined with DFT calculations, we found that the transition at 20 K is rooted in the unique electronic structure of TaSb$_{2}$, which has a shallow shoulder pocket in the vicinity of the main hole pocket. Temperature change first drives the Fermi energy across the band top of the shoulder pocket, leading to the first Lifshitz transition at 20 K. Further temperature increase gradually move the Fermi energy to the band top of the main hole pocket, where the second Lifshitz transition occurs due to the vanishing of hole Fermi surface.

\begin{figure*}[!thb]
\begin{center}
\includegraphics[width=7in]{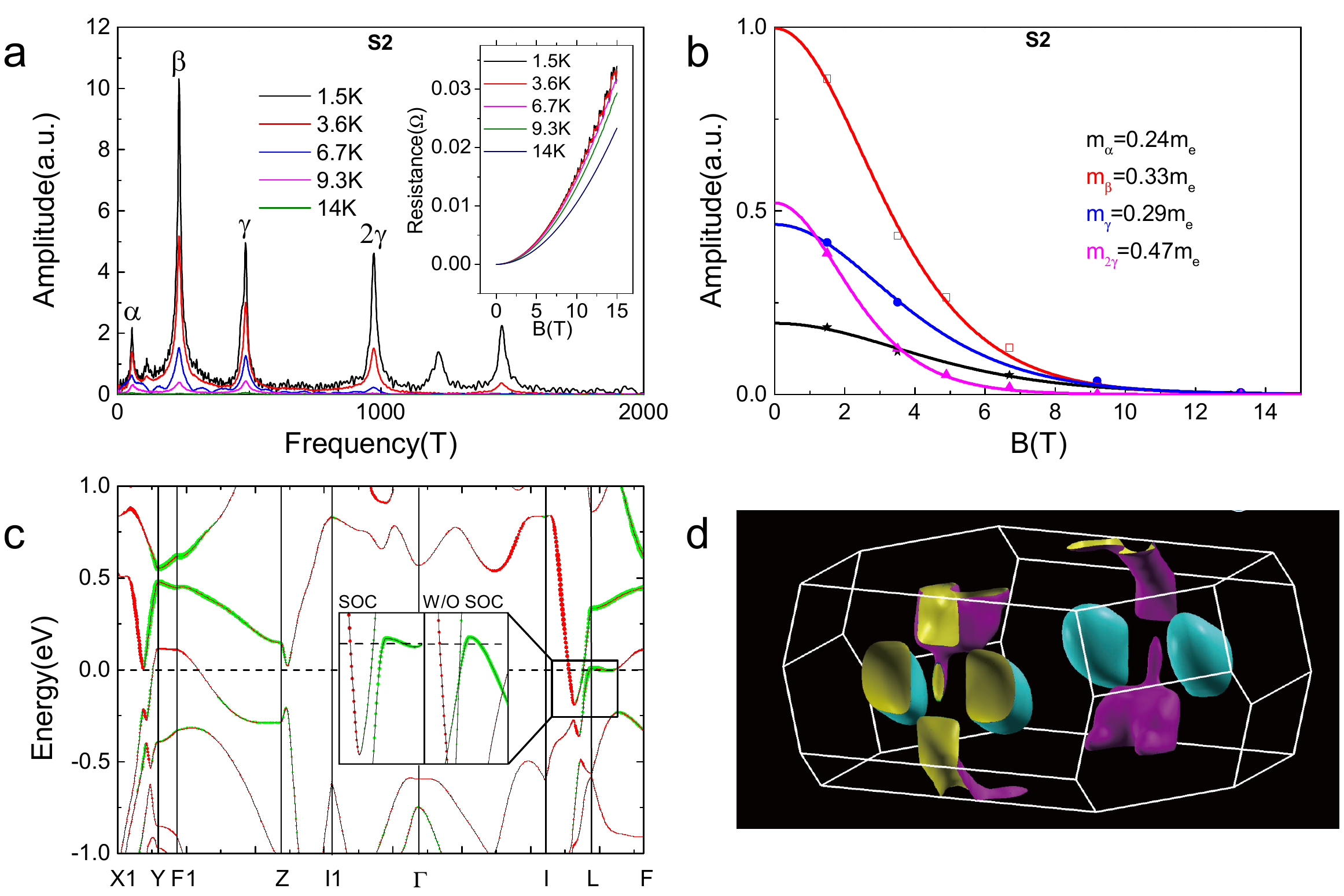}
\end{center}
\caption{\label{Figure4} Unique electronic structure of TaSb$_{2}$. (a) T-dependent SdH oscillations, showing three main FFT frequencies of $\alpha$, $\beta$ and $\gamma$  respectively. The usually high intensity of the second harmonic $2\gamma$ peak is due to magnetic breakdown (see SI). (b) The effective mass of the three pockets. (c) Energy band structure of TaSb$_{2}$ calculated by DFT.  Red and green dots are the projection of Ta-$5d_{xy}$ and Ta-$5d_{x^{2}+y^{2}}$, respectively. The main hole pocket forms a unique shoulder structure along the $F-L$ line, which corresponds to the low frequency $\alpha$ peak in SdH. (d) 3D Fermi surface of TaSb$_{2}$ at 1.5 K, showing the coexistence of compensated $e$-$h$ pockets, and one small hole pocket corresponding to the shoulder structure.}
\end{figure*}

As shown in Figure \ref{Figure4}, the SdH oscillations of TaAs$_{2}$ are characterized by three prominent frequencies of $\alpha=55$ T, $\beta=234$ T, and $\gamma=487$ T, respectively. The results have been confirmed by the magnetic susceptibility oscillations, \textit{i.e.} the de Haas-van Alphen effect (see SI). By fitting the Lifshitz-Kosevich (LK) formula for 3D systems to T-dependent FFT amplitudes \cite{Zeeman_PRB_Balicas}, we can extract the effective mass $m^{\ast}$ for all three pockets, which are $m^{\ast}_{\alpha}=0.24m_{e}$, $m^{\ast}_{\beta}=0.33m_{e}$ and $m^{\ast}_{\gamma}=0.29m_{e}$, respectively (Fig. \ref{Figure4}b). By comparing the FFT peaks with the DFT results, we assign $\beta$ to the $e$ pocket and $\gamma$ to the $h$ pocket, respectively.  At the first glance, the large difference between $\beta$ and $\gamma$ is contradicting with the claim of ideal $e$-$h$ compensation in TaSb$_{2}$. However, DFT indicates the existence of two electron pockets along the $I-L$ lines, while there are only one large hole pocket enclosing the symmetry points of $F_{1}$, $Y$ and $F$  in the first Brillouin zone (Fig. \ref{Figure4}c and Fig. \ref{Figure4}d). Uniquely, each main hole pocket has a much smaller shoulder pocket, which is located along the $F-L$ line and has its band top very close to the Fermi surface ($\sim$9 meV). DFT calculations indicate such shallow shoulder pocket is created by the SOC induced gap opening of the crossed bands of Ta-$5d_{xy}$ and Ta-$5d_{x^{2}+y^{2}}$. In the absence of SOC, there are two band crossing points along the $I-L$ direction (Inset of Fig. \ref{Figure4}c). Unlike the previously reported TIs \cite{TI_RMP10_Kane}, these two band crossing points in TaSb$_{2}$ are inequivalent in energy, which leads to the formation of the unconventional shoulder structure once SOC is turned on. As shown in the right panel of the inset, significant bandgap of 0.45 eV is opened by SOC at both crossing points, making TaSb$_{2}$ a topological semimetal with the topological invariant $Z_{2}$=[0;(111)]. Fig. 4d shows the three-dimensional (3D) Fermi surface of TaSb$_{2}$ at 1.5 K, which visualizes the coexistence of resonant-compensated $e$-$h$ pockets and the $h$ shoulder pocket in between two $e$ pockets. Consequently, \textit{T} increase first leads to the disappearance of the shoulder pocket, which corresponds to the first Lifshitz transition at 20 K (See the schematic in SI). The next topological phase transition is the same type as reported in WTe$_{2}$, which is at the temperature when the Fermi level touching the band top of the main hole pocket (See SI).

\section{Discussion}
We now discuss the possible mechanisms for the unprecedented XMR of TaSb$_{2}$. First, by selecting different FFT windows of the SdH oscillations, we did not see tangible frequency shift in all three peaks. Moreover, the $B^{m}$ growth in XMR starts  at 0.1 T and retains the same $m$ up to 15 T. Thus, we may conclude that the Zeeman effect only contributes marginally to the XMR. Recently, Tafti \textit{et al} suggest that the XMR in LaSb is correlated to the topological surface states and the breaking of time reversal symmetry by the external field \cite{LaSb_Cava_XMR}. TaSb$_{2}$ indeed shares some common features with LaSb in the existence of a TI gap in the bulk and the coexistence of trivial $e$ and $h$ pockets. However, considering the quite large trivial pockets ($n_{e},n_{h} > 10^{19}$ cm$^{-3}$) and the sensitivity of XMR to RRR, it is unlikely that the topological surface state is dominating the experimental observations. 

Our results again testify the importance of resonant $e$-$h$ compensation in the extraordinary XMR phenomena of emergent topological semimetals. Noteworthy, the fast degradation of MR above  20 K is coincident with the disappearance of the small shoulder pocket, despite that TaSb$_{2}$ remains well compensated as indicated by the Nernst results. Considering that the shoulder pocket contributes only 1\% extra holes to the overall hole density (See SI), there may be new mechanism, such as Fermi surface topology, playing a critical role in the XMR phenomena of TaSb$_{2}$. Indeed, the Fermi surface of the hole pocket in TaSb$_{2}$ is rather unconventional, showing not only the shallow shoulder pocket but also a double-saddleback structure near the $F_{1}$ points (See SI). This will require further experimental investigations and theoretical modelling.

In summary, the TaSb$_{2}$ family represent a unique class of topological semimetals, which have the Fermi level well above the SOC-induced TI bandgap and the inverted bands become parts of the Fermi surface. It would be interesting to investigate and compare the XMR and topological phase transition phenomena in other TaSb$_{2}$-family compounds, such as TaAs$_{2}$ and NbAs$_{2}$. For such resonant compensated semimetals, pressure may also lead to the suppression of hole pockets and the emergence of superconductivity, as reported in WTe$_{2}$ \cite{SC_WTe2_NatureComm,SC02_WTe2_NatureComm}.

\begin{acknowledgments}
This work was supported by the National Basic Research Program of China (Grant Nos. 2014CB92103 and 2012CB927404), the National Science Foundation of China (Grant Nos. 11190023, U1332209, 11374009, 61574123 and 11574264), MOE of China (Grant No. 2015KF07), and the Fundamental Research Funds for the Central Universities of China. Y.Zheng acknowledges the start funding support from the 1000 Youth Talent Program.
\end{acknowledgments}


\begin{thebibliography}{34}%
\makeatletter
\providecommand \@ifxundefined [1]{%
 \@ifx{#1\undefined}
}%
\providecommand \@ifnum [1]{%
 \ifnum #1\expandafter \@firstoftwo
 \else \expandafter \@secondoftwo
 \fi
}%
\providecommand \@ifx [1]{%
 \ifx #1\expandafter \@firstoftwo
 \else \expandafter \@secondoftwo
 \fi
}%
\providecommand \natexlab [1]{#1}%
\providecommand \enquote  [1]{``#1''}%
\providecommand \bibnamefont  [1]{#1}%
\providecommand \bibfnamefont [1]{#1}%
\providecommand \citenamefont [1]{#1}%
\providecommand \href@noop [0]{\@secondoftwo}%
\providecommand \href [0]{\begingroup \@sanitize@url \@href}%
\providecommand \@href[1]{\@@startlink{#1}\@@href}%
\providecommand \@@href[1]{\endgroup#1\@@endlink}%
\providecommand \@sanitize@url [0]{\catcode `\\12\catcode `\$12\catcode
  `\&12\catcode `\#12\catcode `\^12\catcode `\_12\catcode `\%12\relax}%
\providecommand \@@startlink[1]{}%
\providecommand \@@endlink[0]{}%
\providecommand \url  [0]{\begingroup\@sanitize@url \@url }%
\providecommand \@url [1]{\endgroup\@href {#1}{\urlprefix }}%
\providecommand \urlprefix  [0]{URL }%
\providecommand \Eprint [0]{\href }%
\providecommand \doibase [0]{http://dx.doi.org/}%
\providecommand \selectlanguage [0]{\@gobble}%
\providecommand \bibinfo  [0]{\@secondoftwo}%
\providecommand \bibfield  [0]{\@secondoftwo}%
\providecommand \translation [1]{[#1]}%
\providecommand \BibitemOpen [0]{}%
\providecommand \bibitemStop [0]{}%
\providecommand \bibitemNoStop [0]{.\EOS\space}%
\providecommand \EOS [0]{\spacefactor3000\relax}%
\providecommand \BibitemShut  [1]{\csname bibitem#1\endcsname}%
\let\auto@bib@innerbib\@empty
\bibitem [{\citenamefont {Kane}\ and\ \citenamefont
  {Mele}(2005)}]{TI_PRL_Kane}%
  \BibitemOpen
  \bibfield  {author} {\bibinfo {author} {\bibfnamefont {C.~L.}\ \bibnamefont
  {Kane}}\ and\ \bibinfo {author} {\bibfnamefont {E.~J.}\ \bibnamefont
  {Mele}},\ }\bibfield  {title} {\enquote {\bibinfo {title} {Z$_{2}$
  topological order and the quantum spin hall effect},}\ }\href@noop {}
  {\bibfield  {journal} {\bibinfo  {journal} {Phys. Rev. Lett.}\ }\textbf
  {\bibinfo {volume} {95}},\ \bibinfo {pages} {146802} (\bibinfo {year}
  {2005})}\BibitemShut {NoStop}%
\bibitem [{\citenamefont {Bernevig}\ \emph {et~al.}(2006)\citenamefont
  {Bernevig}, \citenamefont {Hughes},\ and\ \citenamefont
  {Zhang}}]{TI_Science06_ZSC}%
  \BibitemOpen
  \bibfield  {author} {\bibinfo {author} {\bibfnamefont {B.~A.}\ \bibnamefont
  {Bernevig}}, \bibinfo {author} {\bibfnamefont {T.~L.}\ \bibnamefont
  {Hughes}}, \ and\ \bibinfo {author} {\bibfnamefont {S.~C.}\ \bibnamefont
  {Zhang}},\ }\bibfield  {title} {\enquote {\bibinfo {title} {Quantum spin hall
  effect and topological phase transition in {H}g{T}e quantum wells},}\
  }\href@noop {} {\bibfield  {journal} {\bibinfo  {journal} {Science}\ }\textbf
  {\bibinfo {volume} {314}},\ \bibinfo {pages} {1757--1761} (\bibinfo {year}
  {2006})}\BibitemShut {NoStop}%
\bibitem [{\citenamefont {Hasan}\ and\ \citenamefont
  {Kane}(2010)}]{TI_RMP10_Kane}%
  \BibitemOpen
  \bibfield  {author} {\bibinfo {author} {\bibfnamefont {M.~Z.}\ \bibnamefont
  {Hasan}}\ and\ \bibinfo {author} {\bibfnamefont {C.~L.}\ \bibnamefont
  {Kane}},\ }\bibfield  {title} {\enquote {\bibinfo {title} {Colloquium:
  Topological insulators},}\ }\href@noop {} {\bibfield  {journal} {\bibinfo
  {journal} {Rev. Mod. Phys.}\ }\textbf {\bibinfo {volume} {82}},\ \bibinfo
  {pages} {3405--3067} (\bibinfo {year} {2010})}\BibitemShut {NoStop}%
\bibitem [{\citenamefont {Young}\ and\ \citenamefont
  {et~al}(2012)}]{DSM_PRLtheory_Kane}%
  \BibitemOpen
  \bibfield  {author} {\bibinfo {author} {\bibfnamefont {S.~M.}\ \bibnamefont
  {Young}}\ and\ \bibinfo {author} {\bibnamefont {et~al}},\ }\bibfield  {title}
  {\enquote {\bibinfo {title} {Dirac semimetal in three dimensions},}\
  }\href@noop {} {\bibfield  {journal} {\bibinfo  {journal} {Phys. Rev. Lett.}\
  }\textbf {\bibinfo {volume} {108}},\ \bibinfo {pages} {140405} (\bibinfo
  {year} {2012})}\BibitemShut {NoStop}%
\bibitem [{\citenamefont {Wang}\ \emph {et~al.}(2013)\citenamefont {Wang},
  \citenamefont {Weng}, \citenamefont {Q.~Wu},\ and\ \citenamefont
  {Fang}}]{Cd3As2_PRBtheory_LMR}%
  \BibitemOpen
  \bibfield  {author} {\bibinfo {author} {\bibfnamefont {Z.}~\bibnamefont
  {Wang}}, \bibinfo {author} {\bibfnamefont {H.}~\bibnamefont {Weng}}, \bibinfo
  {author} {\bibfnamefont {X.~Dai}\ \bibnamefont {Q.~Wu}}, \ and\ \bibinfo
  {author} {\bibfnamefont {Z.}~\bibnamefont {Fang}},\ }\bibfield  {title}
  {\enquote {\bibinfo {title} {Three-dimensional dirac semimetal and quantum
  transport in {C}d$_{3}${A}s$_{2}$},}\ }\href@noop {} {\bibfield  {journal}
  {\bibinfo  {journal} {Phys. Rev. B}\ }\textbf {\bibinfo {volume} {88}},\
  \bibinfo {pages} {125427} (\bibinfo {year} {2013})}\BibitemShut {NoStop}%
\bibitem [{\citenamefont {Tian}\ \emph {et~al.}(2015)\citenamefont {Tian},
  \citenamefont {Gibson}, \citenamefont {Ali}, \citenamefont {Liu},
  \citenamefont {Cava},\ and\ \citenamefont {Ong}}]{Cd3As2_NPOng_NatMat15}%
  \BibitemOpen
  \bibfield  {author} {\bibinfo {author} {\bibfnamefont {L.}~\bibnamefont
  {Tian}}, \bibinfo {author} {\bibfnamefont {Q.}~\bibnamefont {Gibson}},
  \bibinfo {author} {\bibfnamefont {M.~N.}\ \bibnamefont {Ali}}, \bibinfo
  {author} {\bibfnamefont {M.}~\bibnamefont {Liu}}, \bibinfo {author}
  {\bibfnamefont {R.~J.}\ \bibnamefont {Cava}}, \ and\ \bibinfo {author}
  {\bibfnamefont {N.~P.}\ \bibnamefont {Ong}},\ }\bibfield  {title} {\enquote
  {\bibinfo {title} {Ultrahigh mobility and giant magnetoresistance in the
  {D}irac semimetal {C}d$_{3}${A}s$_{2}$},}\ }\href@noop {} {\bibfield
  {journal} {\bibinfo  {journal} {Nature Mater.}\ }\textbf {\bibinfo {volume}
  {14}},\ \bibinfo {pages} {280--284} (\bibinfo {year} {2015})}\BibitemShut
  {NoStop}%
\bibitem [{\citenamefont {Wan}\ \emph {et~al.}(2011)\citenamefont {Wan},
  \citenamefont {Turner}, \citenamefont {Vishwanath},\ and\ \citenamefont
  {Savrasov}}]{WSMWanXG_PRB}%
  \BibitemOpen
  \bibfield  {author} {\bibinfo {author} {\bibfnamefont {X.~G.}\ \bibnamefont
  {Wan}}, \bibinfo {author} {\bibfnamefont {A.~M.}\ \bibnamefont {Turner}},
  \bibinfo {author} {\bibfnamefont {A.}~\bibnamefont {Vishwanath}}, \ and\
  \bibinfo {author} {\bibfnamefont {S.~Y.}\ \bibnamefont {Savrasov}},\
  }\bibfield  {title} {\enquote {\bibinfo {title} {Topological semimetal and
  {F}ermi-arc surface states in the electronic structure of pyrochlore
  iridates},}\ }\href@noop {} {\bibfield  {journal} {\bibinfo  {journal} {Phys.
  Rev. B}\ }\textbf {\bibinfo {volume} {83}},\ \bibinfo {pages} {205101}
  (\bibinfo {year} {2011})}\BibitemShut {NoStop}%
\bibitem [{\citenamefont {Weng}\ \emph {et~al.}(2015)\citenamefont {Weng},
  \citenamefont {Fang}, \citenamefont {Fang}, \citenamefont {Bernevig},\ and\
  \citenamefont {Dai}}]{WSMDaiX_PRX}%
  \BibitemOpen
  \bibfield  {author} {\bibinfo {author} {\bibfnamefont {H.}~\bibnamefont
  {Weng}}, \bibinfo {author} {\bibfnamefont {C.}~\bibnamefont {Fang}}, \bibinfo
  {author} {\bibfnamefont {Z.}~\bibnamefont {Fang}}, \bibinfo {author}
  {\bibfnamefont {B.~A.}\ \bibnamefont {Bernevig}}, \ and\ \bibinfo {author}
  {\bibfnamefont {X.}~\bibnamefont {Dai}},\ }\bibfield  {title} {\enquote
  {\bibinfo {title} {Weyl semimetal phase in noncentrosymmetric
  transition-metal monophosphides},}\ }\href@noop {} {\bibfield  {journal}
  {\bibinfo  {journal} {Phys. Rev. X}\ }\textbf {\bibinfo {volume} {5}},\
  \bibinfo {pages} {011029} (\bibinfo {year} {2015})}\BibitemShut {NoStop}%
\bibitem [{\citenamefont {Huang}\ and\ \citenamefont
  {et~al.}(2014)}]{NCHasan_WSMTheory}%
  \BibitemOpen
  \bibfield  {author} {\bibinfo {author} {\bibfnamefont {S.}~\bibnamefont
  {Huang}}\ and\ \bibinfo {author} {\bibnamefont {et~al.}},\ }\bibfield
  {title} {\enquote {\bibinfo {title} {An inversion breaking {W}eyl semimetal
  state in the {T}a{A}s material class},}\ }\href@noop {} {\bibfield  {journal}
  {\bibinfo  {journal} {Nature Commun.}\ }\textbf {\bibinfo {volume} {6}},\
  \bibinfo {pages} {7373} (\bibinfo {year} {2014})}\BibitemShut {NoStop}%
\bibitem [{\citenamefont {Bian}\ and\ \citenamefont
  {et~al.}(2015)}]{PbTaSe2_Hasan_Nodal}%
  \BibitemOpen
  \bibfield  {author} {\bibinfo {author} {\bibfnamefont {G.}~\bibnamefont
  {Bian}}\ and\ \bibinfo {author} {\bibnamefont {et~al.}},\ }\bibfield  {title}
  {\enquote {\bibinfo {title} {Topological nodal-line fermions in the
  non-centrosymmetric superconductor compound {P}b{T}a{S}e$_{2}$},}\
  }\href@noop {} {\bibfield  {journal} {\bibinfo  {journal} {arXiv:1505.03069}\
  } (\bibinfo {year} {2015})}\BibitemShut {NoStop}%
\bibitem [{\citenamefont {{Witczak-Krempa}}\ \emph {et~al.}(2014)\citenamefont
  {{Witczak-Krempa}}, \citenamefont {Chen}, \citenamefont {Kim},\ and\
  \citenamefont {Balents}}]{QuatumPhases_Balents_AR}%
  \BibitemOpen
  \bibfield  {author} {\bibinfo {author} {\bibfnamefont {W.}~\bibnamefont
  {{Witczak-Krempa}}}, \bibinfo {author} {\bibfnamefont {G.}~\bibnamefont
  {Chen}}, \bibinfo {author} {\bibfnamefont {Y.~B.}\ \bibnamefont {Kim}}, \
  and\ \bibinfo {author} {\bibfnamefont {L.}~\bibnamefont {Balents}},\
  }\bibfield  {title} {\enquote {\bibinfo {title} {Correlated quantum phenomena
  in the strong spin-orbit regime},}\ }\href@noop {} {\bibfield  {journal}
  {\bibinfo  {journal} {Annu. Rev. Condens. Matter Phys.}\ }\textbf {\bibinfo
  {volume} {5}} (\bibinfo {year} {2014})}\BibitemShut {NoStop}%
\bibitem [{\citenamefont {Narayanan}\ and\ \citenamefont
  {et~al}(2015)}]{Cd3As2_PRLXMR_Coldea}%
  \BibitemOpen
  \bibfield  {author} {\bibinfo {author} {\bibfnamefont {A.}~\bibnamefont
  {Narayanan}}\ and\ \bibinfo {author} {\bibnamefont {et~al}},\ }\bibfield
  {title} {\enquote {\bibinfo {title} {Linear magnetoresistance caused by
  mobility fluctuations in n-doped {C}d$_{3}${A}s$_{2}$},}\ }\href@noop {}
  {\bibfield  {journal} {\bibinfo  {journal} {Phys. Rev. Lett.}\ }\textbf
  {\bibinfo {volume} {114}},\ \bibinfo {pages} {117201} (\bibinfo {year}
  {2015})}\BibitemShut {NoStop}%
\bibitem [{\citenamefont {Zhang}\ \emph {et~al.}(2015)\citenamefont {Zhang},
  \citenamefont {Yuan}, \citenamefont {Xu}, \citenamefont {Lin}, \citenamefont
  {Tong}, \citenamefont {Hasan}, \citenamefont {Wang}, \citenamefont {Zhang},\
  and\ \citenamefont {Jia}}]{TaAs_arXiv_Jia}%
  \BibitemOpen
  \bibfield  {author} {\bibinfo {author} {\bibfnamefont {C.}~\bibnamefont
  {Zhang}}, \bibinfo {author} {\bibfnamefont {Z.}~\bibnamefont {Yuan}},
  \bibinfo {author} {\bibfnamefont {S.}~\bibnamefont {Xu}}, \bibinfo {author}
  {\bibfnamefont {Z.}~\bibnamefont {Lin}}, \bibinfo {author} {\bibfnamefont
  {B.}~\bibnamefont {Tong}}, \bibinfo {author} {\bibfnamefont {M.~Z.}\
  \bibnamefont {Hasan}}, \bibinfo {author} {\bibfnamefont {J.}~\bibnamefont
  {Wang}}, \bibinfo {author} {\bibfnamefont {C.}~\bibnamefont {Zhang}}, \ and\
  \bibinfo {author} {\bibfnamefont {S.}~\bibnamefont {Jia}},\ }\bibfield
  {title} {\enquote {\bibinfo {title} {Tantalum monoarsenide: an exotic
  compensated semimetal},}\ }\href@noop {} {\bibfield  {journal} {\bibinfo
  {journal} {arXiv:1502.00251}\ } (\bibinfo {year} {2015})}\BibitemShut
  {NoStop}%
\bibitem [{\citenamefont {Huang}\ and\ \citenamefont
  {et~al.}(2015)}]{TaAs_PRX_NMR}%
  \BibitemOpen
  \bibfield  {author} {\bibinfo {author} {\bibfnamefont {X.}~\bibnamefont
  {Huang}}\ and\ \bibinfo {author} {\bibnamefont {et~al.}},\ }\bibfield
  {title} {\enquote {\bibinfo {title} {Observation of the chiral anomaly
  induced negative magneto-resistance in 3{D} {W}eyl semi-metal {T}a{A}s},}\
  }\href@noop {} {\bibfield  {journal} {\bibinfo  {journal} {Phys. Rev. X}\
  }\textbf {\bibinfo {volume} {5}},\ \bibinfo {pages} {031023} (\bibinfo {year}
  {2015})}\BibitemShut {NoStop}%
\bibitem [{\citenamefont {Ghimire}\ and\ \citenamefont
  {et~al.}(2015)}]{NbAs_jpcm}%
  \BibitemOpen
  \bibfield  {author} {\bibinfo {author} {\bibfnamefont {N.~J.}\ \bibnamefont
  {Ghimire}}\ and\ \bibinfo {author} {\bibnamefont {et~al.}},\ }\bibfield
  {title} {\enquote {\bibinfo {title} {Magnetotransport of single crystalline
  {N}b{A}s},}\ }\href@noop {} {\bibfield  {journal} {\bibinfo  {journal} {J.
  Phys.: Condens. Matter}\ }\textbf {\bibinfo {volume} {27}},\ \bibinfo {pages}
  {152201} (\bibinfo {year} {2015})}\BibitemShut {NoStop}%
\bibitem [{\citenamefont {Shekhar}\ and\ \citenamefont
  {et~al.}(2015)}]{NbP_NaturePhy}%
  \BibitemOpen
  \bibfield  {author} {\bibinfo {author} {\bibfnamefont {C.}~\bibnamefont
  {Shekhar}}\ and\ \bibinfo {author} {\bibnamefont {et~al.}},\ }\bibfield
  {title} {\enquote {\bibinfo {title} {Extremely large magnetoresistance and
  ultrahigh mobility in the topological {W}eyl semimetal {N}b{P}},}\
  }\href@noop {} {\bibfield  {journal} {\bibinfo  {journal} {Nature Phys.}\
  }\textbf {\bibinfo {volume} {11}},\ \bibinfo {pages} {645} (\bibinfo {year}
  {2015})}\BibitemShut {NoStop}%
\bibitem [{\citenamefont {Wang}\ \emph {et~al.}(2015)\citenamefont {Wang},
  \citenamefont {Zheng}, \citenamefont {Shen}, \citenamefont {Zhou},
  \citenamefont {Yang}, \citenamefont {Li}, \citenamefont {Feng},\ and\
  \citenamefont {Xu}}]{NbP_arXivWZ}%
  \BibitemOpen
  \bibfield  {author} {\bibinfo {author} {\bibfnamefont {Z.}~\bibnamefont
  {Wang}}, \bibinfo {author} {\bibfnamefont {Y.}~\bibnamefont {Zheng}},
  \bibinfo {author} {\bibfnamefont {Z.~X.}\ \bibnamefont {Shen}}, \bibinfo
  {author} {\bibfnamefont {Y.}~\bibnamefont {Zhou}}, \bibinfo {author}
  {\bibfnamefont {X.~J.}\ \bibnamefont {Yang}}, \bibinfo {author}
  {\bibfnamefont {Y.~P.}\ \bibnamefont {Li}}, \bibinfo {author} {\bibfnamefont
  {C.~M.}\ \bibnamefont {Feng}}, \ and\ \bibinfo {author} {\bibfnamefont
  {Z.~A.}\ \bibnamefont {Xu}},\ }\bibfield  {title} {\enquote {\bibinfo {title}
  {Helicity protected ultrahigh mobility {W}eyl fermions in {N}b{P}},}\
  }\href@noop {} {\bibfield  {journal} {\bibinfo  {journal} {arXiv:1506.00924}\
  } (\bibinfo {year} {2015})}\BibitemShut {NoStop}%
\bibitem [{\citenamefont {Ali}\ and\ \citenamefont
  {et~al.}(2014)}]{XMR_WTe2NPOng_Nature14}%
  \BibitemOpen
  \bibfield  {author} {\bibinfo {author} {\bibfnamefont {M.~N.}\ \bibnamefont
  {Ali}}\ and\ \bibinfo {author} {\bibnamefont {et~al.}},\ }\bibfield  {title}
  {\enquote {\bibinfo {title} {Large, non-saturating magnetoresistance in
  {W}{T}e$_{2}$},}\ }\href@noop {} {\bibfield  {journal} {\bibinfo  {journal}
  {Nature}\ }\textbf {\bibinfo {volume} {514}},\ \bibinfo {pages} {205}
  (\bibinfo {year} {2014})}\BibitemShut {NoStop}%
\bibitem [{\citenamefont {Pletikosi{\'{c}}}\ and\ \citenamefont
  {et~al.}(2014)}]{Aupes_PRL_Valla}%
  \BibitemOpen
  \bibfield  {author} {\bibinfo {author} {\bibfnamefont {I.}~\bibnamefont
  {Pletikosi{\'{c}}}}\ and\ \bibinfo {author} {\bibnamefont {et~al.}},\
  }\bibfield  {title} {\enquote {\bibinfo {title} {Electronic structure basis
  for the extraordinary magnetoresistance in {W}{T}e$_{2}$},}\ }\href@noop {}
  {\bibfield  {journal} {\bibinfo  {journal} {Phys. Rev. Lett.}\ }\textbf
  {\bibinfo {volume} {113}},\ \bibinfo {pages} {216601} (\bibinfo {year}
  {2014})}\BibitemShut {NoStop}%
\bibitem [{\citenamefont {Zhu}\ and\ \citenamefont
  {et~al.}(2015)}]{FS_PRL_ZhuZW}%
  \BibitemOpen
  \bibfield  {author} {\bibinfo {author} {\bibfnamefont {Z.~W.}\ \bibnamefont
  {Zhu}}\ and\ \bibinfo {author} {\bibnamefont {et~al.}},\ }\bibfield  {title}
  {\enquote {\bibinfo {title} {Quantum oscillations, thermoelectric
  coefficients, and the {F}ermi surface of semimetallic {W}{T}e$_{2}$},}\
  }\href@noop {} {\bibfield  {journal} {\bibinfo  {journal} {Phys. Rev. Lett.}\
  }\textbf {\bibinfo {volume} {114}},\ \bibinfo {pages} {176601} (\bibinfo
  {year} {2015})}\BibitemShut {NoStop}%
\bibitem [{\citenamefont {Jiang}\ and\ \citenamefont
  {et~al.}(2015)}]{SOC_Aupes_PRL_FengDL}%
  \BibitemOpen
  \bibfield  {author} {\bibinfo {author} {\bibfnamefont {J.}~\bibnamefont
  {Jiang}}\ and\ \bibinfo {author} {\bibnamefont {et~al.}},\ }\bibfield
  {title} {\enquote {\bibinfo {title} {Signature of strong spin-orbital
  coupling in the large nonsaturating magnetoresistance material
  {W}{T}e$_{2}$},}\ }\href@noop {} {\bibfield  {journal} {\bibinfo  {journal}
  {Phys. Rev. Lett.}\ }\textbf {\bibinfo {volume} {115}},\ \bibinfo {pages}
  {166601} (\bibinfo {year} {2015})}\BibitemShut {NoStop}%
\bibitem [{\citenamefont {Rhodes}\ and\ \citenamefont
  {et~al.}(2015)}]{Zeeman_PRB_Balicas}%
  \BibitemOpen
  \bibfield  {author} {\bibinfo {author} {\bibfnamefont {D.}~\bibnamefont
  {Rhodes}}\ and\ \bibinfo {author} {\bibnamefont {et~al.}},\ }\bibfield
  {title} {\enquote {\bibinfo {title} {Role of spin-orbit coupling and
  evolution of the electronic structure of {W}{T}e$_{2}$ under an external
  magnetic field},}\ }\href@noop {} {\bibfield  {journal} {\bibinfo  {journal}
  {Phys. Rev. B}\ }\textbf {\bibinfo {volume} {92}},\ \bibinfo {pages} {125152}
  (\bibinfo {year} {2015})}\BibitemShut {NoStop}%
\bibitem [{\citenamefont {Tafti}\ and\ \citenamefont
  {et~al.}(2015)}]{LaSb_Cava_XMR}%
  \BibitemOpen
  \bibfield  {author} {\bibinfo {author} {\bibfnamefont {F.~F.}\ \bibnamefont
  {Tafti}}\ and\ \bibinfo {author} {\bibnamefont {et~al.}},\ }\bibfield
  {title} {\enquote {\bibinfo {title} {Consequences of breaking time reversal
  symmetry in {L}a{S}b: a resistivity plateau and extreme magnetoresistance},}\
  }\href@noop {} {\bibfield  {journal} {\bibinfo  {journal} {arXiv:1510.06931}\
  } (\bibinfo {year} {2015})}\BibitemShut {NoStop}%
\bibitem [{\citenamefont {Soluyanov}\ and\ \citenamefont
  {et~al}(2015)}]{WSMIIWTe2_Bernevig_Nature15}%
  \BibitemOpen
  \bibfield  {author} {\bibinfo {author} {\bibfnamefont {A.~A.}\ \bibnamefont
  {Soluyanov}}\ and\ \bibinfo {author} {\bibnamefont {et~al}},\ }\bibfield
  {title} {\enquote {\bibinfo {title} {Type-{II} {W}eyl semimetals},}\
  }\href@noop {} {\bibfield  {journal} {\bibinfo  {journal} {Nature}\ }\textbf
  {\bibinfo {volume} {527}},\ \bibinfo {pages} {495--498} (\bibinfo {year}
  {2015})}\BibitemShut {NoStop}%
\bibitem [{\citenamefont {Zeng}\ and\ \citenamefont
  {et~al.}(2015)}]{LaSbTI_LinHsin_arXiv}%
  \BibitemOpen
  \bibfield  {author} {\bibinfo {author} {\bibfnamefont {M.~G.}\ \bibnamefont
  {Zeng}}\ and\ \bibinfo {author} {\bibnamefont {et~al.}},\ }\bibfield  {title}
  {\enquote {\bibinfo {title} {Topological semimetals and topological
  insulators in rare earth monopnictides},}\ }\href@noop {} {\bibfield
  {journal} {\bibinfo  {journal} {arXiv:1504.03492}\ } (\bibinfo {year}
  {2015})}\BibitemShut {NoStop}%
\bibitem [{\citenamefont {Blochl}(1994)}]{DFT_Blochl_PRB94}%
  \BibitemOpen
  \bibfield  {author} {\bibinfo {author} {\bibfnamefont {P.~E.}\ \bibnamefont
  {Blochl}},\ }\bibfield  {title} {\enquote {\bibinfo {title} {Projector
  augmented-wave method},}\ }\href@noop {} {\bibfield  {journal} {\bibinfo
  {journal} {Phys. Rev. B}\ }\textbf {\bibinfo {volume} {50}} (\bibinfo {year}
  {1994})}\BibitemShut {NoStop}%
\bibitem [{\citenamefont {Kresse}\ and\ \citenamefont
  {Hafner}(1993)}]{VASP_Kresse_PRB93}%
  \BibitemOpen
  \bibfield  {author} {\bibinfo {author} {\bibfnamefont {G.}~\bibnamefont
  {Kresse}}\ and\ \bibinfo {author} {\bibfnamefont {J.}~\bibnamefont
  {Hafner}},\ }\bibfield  {title} {\enquote {\bibinfo {title} {Ab initio
  molecular-dynamics for liquid-metals},}\ }\href@noop {} {\bibfield  {journal}
  {\bibinfo  {journal} {Phys. Rev. B}\ }\textbf {\bibinfo {volume} {47}}
  (\bibinfo {year} {1993})}\BibitemShut {NoStop}%
\bibitem [{\citenamefont {Kresse}\ and\ \citenamefont
  {Furthmuller}(1996)}]{VASP_Kresse_PRB96}%
  \BibitemOpen
  \bibfield  {author} {\bibinfo {author} {\bibfnamefont {G.}~\bibnamefont
  {Kresse}}\ and\ \bibinfo {author} {\bibfnamefont {J.}~\bibnamefont
  {Furthmuller}},\ }\bibfield  {title} {\enquote {\bibinfo {title} {Efficient
  iterative schemes for ab initio total-energy calculations using a plane-wave
  basis set},}\ }\href@noop {} {\bibfield  {journal} {\bibinfo  {journal}
  {Phys. Rev. B}\ }\textbf {\bibinfo {volume} {54}} (\bibinfo {year}
  {1996})}\BibitemShut {NoStop}%
\bibitem [{\citenamefont {Perdew}\ \emph {et~al.}(1996)\citenamefont {Perdew},
  \citenamefont {Burke},\ and\ \citenamefont {Ernzerhof}}]{GGA_Perdew_PRL96}%
  \BibitemOpen
  \bibfield  {author} {\bibinfo {author} {\bibfnamefont {J.~P.}\ \bibnamefont
  {Perdew}}, \bibinfo {author} {\bibfnamefont {K.}~\bibnamefont {Burke}}, \
  and\ \bibinfo {author} {\bibfnamefont {M.}~\bibnamefont {Ernzerhof}},\
  }\bibfield  {title} {\enquote {\bibinfo {title} {Generalized gradient
  approximation made simple},}\ }\href@noop {} {\bibfield  {journal} {\bibinfo
  {journal} {Phys. Rev. Lett.}\ }\textbf {\bibinfo {volume} {77}} (\bibinfo
  {year} {1996})}\BibitemShut {NoStop}%
\bibitem [{\citenamefont {Monkhorst}\ and\ \citenamefont
  {Pack}(1996)}]{MPscheme_Monkhorst_PRB76}%
  \BibitemOpen
  \bibfield  {author} {\bibinfo {author} {\bibfnamefont {H.~J.}\ \bibnamefont
  {Monkhorst}}\ and\ \bibinfo {author} {\bibfnamefont {J.~D.}\ \bibnamefont
  {Pack}},\ }\bibfield  {title} {\enquote {\bibinfo {title} {Special points for
  {B}rillouin-zone integrations},}\ }\href@noop {} {\bibfield  {journal}
  {\bibinfo  {journal} {Phys. Rev. B}\ }\textbf {\bibinfo {volume} {13}}
  (\bibinfo {year} {1996})}\BibitemShut {NoStop}%
\bibitem [{\citenamefont {Hurd}(1972)}]{HallinMetal_Colin}%
  \BibitemOpen
  \bibfield  {author} {\bibinfo {author} {\bibfnamefont {Colin~M.}\
  \bibnamefont {Hurd}},\ }\href@noop {} {\emph {\bibinfo {title} {The Hall
  Effect in Metals and Alloys}}}\ (\bibinfo  {publisher} {Cambridge University
  Press, Cambridge},\ \bibinfo {year} {1972})\BibitemShut {NoStop}%
\bibitem [{\citenamefont {Wu}\ and\ \citenamefont
  {et~al.}(2015)}]{Lifshitz_PRL_Kaminski}%
  \BibitemOpen
  \bibfield  {author} {\bibinfo {author} {\bibfnamefont {Y.}~\bibnamefont
  {Wu}}\ and\ \bibinfo {author} {\bibnamefont {et~al.}},\ }\bibfield  {title}
  {\enquote {\bibinfo {title} {Temperature-induced {L}ifshitz transition in
  {W}{T}e$_{2}$},}\ }\href@noop {} {\bibfield  {journal} {\bibinfo  {journal}
  {Phys. Rev. Lett.}\ }\textbf {\bibinfo {volume} {115}},\ \bibinfo {pages}
  {166602} (\bibinfo {year} {2015})}\BibitemShut {NoStop}%
\bibitem [{\citenamefont {Kang}\ and\ \citenamefont
  {et~al.}(2015)}]{SC_WTe2_NatureComm}%
  \BibitemOpen
  \bibfield  {author} {\bibinfo {author} {\bibfnamefont {D.~F.}\ \bibnamefont
  {Kang}}\ and\ \bibinfo {author} {\bibnamefont {et~al.}},\ }\bibfield  {title}
  {\enquote {\bibinfo {title} {Superconductivity emerging from a suppressed
  large magnetoresistant state in tungsten ditelluride},}\ }\href@noop {}
  {\bibfield  {journal} {\bibinfo  {journal} {Nature Comm.}\ }\textbf {\bibinfo
  {volume} {6}},\ \bibinfo {pages} {7804} (\bibinfo {year} {2015})}\BibitemShut
  {NoStop}%
\bibitem [{\citenamefont {Pan}\ and\ \citenamefont
  {et~al.}(2015)}]{SC02_WTe2_NatureComm}%
  \BibitemOpen
  \bibfield  {author} {\bibinfo {author} {\bibfnamefont {X.~C.}\ \bibnamefont
  {Pan}}\ and\ \bibinfo {author} {\bibnamefont {et~al.}},\ }\bibfield  {title}
  {\enquote {\bibinfo {title} {Pressure-driven dome-shaped superconductivity
  and electronic structural evolution in tungsten ditelluride},}\ }\href@noop
  {} {\bibfield  {journal} {\bibinfo  {journal} {Nature Comm.}\ }\textbf
  {\bibinfo {volume} {6}},\ \bibinfo {pages} {7805} (\bibinfo {year}
  {2015})}\BibitemShut {NoStop}%
\end{thebibliography}
\end{document}